\documentclass[
    10pt,
    aps,
    pra,
    reprint,
    a4paper,
    amsfonts, 
    amssymb, 
    amsmath,  
    showkeys,
    superscriptaddress,
]{revtex4-2}
\usepackage[english]{babel}
\usepackage[latin1]{inputenc}
\usepackage[T1]{fontenc}
\usepackage[final]{microtype}

\usepackage{amsthm}
\usepackage{mathtools}
\usepackage{physics}
\usepackage{xcolor}
\usepackage{graphicx}
\usepackage{adjustbox}
\usepackage{placeins}
\usepackage{lipsum}
\usepackage{csquotes}
\usepackage{comment}

\usepackage{siunitx}
\usepackage[
    colorlinks=true,
    allcolors=blue,
    pdftex,
    pdftitle={Article},
    pdfauthor={Author},
]{hyperref}
\usepackage[all]{hypcap}
\usepackage{cleveref}

\Crefname{figure}{Fig.}{Figs.}

\begin{document}
\title{Industrially fabricated single-electron quantum dots in Si/Si-Ge heterostructures}
\thanks{This work has been submitted to the IEEE for possible publication. Copyright may be transferred without notice, after which this version may no longer be accessible.}

\author{Till Huckemann}
    \email{till.huckemann@rwth-aachen.de}% Your name
    \affiliation{JARA Institute for Quantum Information, RWTH Aachen University, 52062 Aachen, Germany}
\author{Pascal Muster}
    \email{pascal.muster@infineon.com}% Your name
    \affiliation{Infineon Technologies Dresden GmbH \& Co. KG, K\"onigsbr\"ucker-Str. 180, 01099 Dresden}
\author{Wolfram Langheinrich}
    \affiliation{Infineon Technologies Dresden GmbH \& Co. KG, K\"onigsbr\"ucker-Str. 180, 01099 Dresden}
\author{Varvara Brackmann}
    \affiliation{Fraunhofer Institute for Photonic Microsystems, Center for Nanoelectronic Technologies, An der Bartlake 5, 01099 Dresden, Germany}
\author{Michael Friedrich}
    \affiliation{Fraunhofer Institute for Photonic Microsystems, Center for Nanoelectronic Technologies, An der Bartlake 5, 01099 Dresden, Germany}
\author{Nikola D. Komeri{\v c}ki}
    \affiliation{Fraunhofer Institute for Applied Solid State Physics IAF, Tullastr. 72, 79108 Freiburg, Germany}
\author{Laura K. Diebel}
    \affiliation{Fakult\"at f\"ur Physik, Universit\"at Regensburg, 93040 Regensburg, Germany}
\author{Verena Stie\ss}
    \affiliation{Fakult\"at f\"ur Physik, Universit\"at Regensburg, 93040 Regensburg, Germany}
\author{Dominique Bougeard}
    \affiliation{Fakult\"at f\"ur Physik, Universit\"at Regensburg, 93040 Regensburg, Germany}
\author{Yuji Yamamoto}
    \affiliation{IHP -- Leibniz-Institut f\"ur innovative Mikroelektronik, 15236 Frankfurt (Oder), Germany}
\author{Felix Reichmann}
    \affiliation{IHP -- Leibniz-Institut f\"ur innovative Mikroelektronik, 15236 Frankfurt (Oder), Germany}
\author{Marvin H. Z\"ollner}
    \affiliation{IHP -- Leibniz-Institut f\"ur innovative Mikroelektronik, 15236 Frankfurt (Oder), Germany}
\author{Claus Dahl}
    \affiliation{Infineon Technologies Dresden GmbH \& Co. KG, K\"onigsbr\"ucker-Str. 180, 01099 Dresden}
\author{Lars R. Schreiber}
    \affiliation{JARA Institute for Quantum Information, RWTH Aachen University, 52062 Aachen, Germany}
    \affiliation{ARQUE Systems GmbH, 52074 Aachen, Germany}
\author{Hendrik Bluhm}
    \affiliation{JARA Institute for Quantum Information, RWTH Aachen University, 52062 Aachen, Germany}
    \affiliation{ARQUE Systems GmbH, 52074 Aachen, Germany}

\date{17.02.2025} % Leave empty to omit a date

\begin{abstract}
This paper reports the compatibility of heterostructure-based spin qubit devices with industrial CMOS technology. It features Si/Si--Ge quantum dot devices fabricated using Infineon's 200\,mm production line within a restricted thermal budget. The devices exhibit state-of-the-art charge sensing, charge noise and valley splitting characteristics, showing that industrial fabrication is not harming the heterostructure quality. These measured parameters are all correlated to spin qubit coherence and qubit gate fidelity.
We describe the single electron device layout, design and its fabrication process using electron beam lithography. The incorporated standard \SI{90}{\nano\meter} back-end of line flow for gate-layer independent contacting and wiring can be scaled up to multiple wiring layers for scalable quantum computing architectures. In addition, we present millikelvin characterization results.
Our work exemplifies the potential of industrial fabrication methods to harness the inherent CMOS-compatibility of the Si/Si--Ge material system, despite being restricted to a reduced thermal budget. It paves the way for advanced quantum processor architectures with high yield and device quality.
\end{abstract}


\maketitle
\section{Introduction}
\label{sec:introduction}

Universal quantum computers with error correction will require orders of magnitude more qubits than available today. Semiconductor qubits offer the perspective to take advantage of very large scale integration (VLSI) technology that can easily achieve such a complexity for classical electronics. Using devices fabricated in academic labs, all required qubit operations \cite{Yoneda2018, Noiri2022, Mills2022}, few-qubit devices \cite{Philips2022, Ha22} and qubit shuttling as a scalability element \cite{Struck2024, Xue2024, deSmets2024} have been demonstrated.  Now, the transfer to industrial fabrication techniques is required to achieve a sufficient yield for proposed quantum processor architectures \cite{veldhorst2017,boter2022,Kunne2024}.
The platforms on which industrial progress has recently been made include fully depleted silicon on insulator (FDSOI) nanowires \cite{Bedecarrats2021}, FinFET devices  \cite{Zwerver2022} and Si/Si--Ge heterostructures \cite{Ziegler2023} using CMOS pilot lines.
While the first two types are compatible with planar CMOS processing techniques, heterostructures impose a restricted thermal budget to avoid excessive inter-diffusion at the interfaces \cite{Klos2024}. It limits the options for oxide formation to temperatures below $\approx \SI{750}{\celsius}$ which raises concerns regarding oxide quality and thus charge noise.
On the other hand, a heterostructure-based quantum well has the advantage of an all-crystalline environment and therefore lower disorder, provided that the dislocation density in the crystalline material is sufficiently low.  As a result, multi-quantum-dot and multi-qubit devices have already been realized \cite{Ha22,Philips2022}.

Besides yield and qubit quality, as recently demonstrated in Si/Si--Ge \cite{Neyens2024,Philips2022} using a high-$\kappa$ gate stack process, the proposed quantum processor architectures also require a Back-End of Line (BEoL) processes for multi-layer wiring schemes.

We report the realization of quantum dot (QD) devices in Si/Si--Ge heterostructures in a \SI{200}{\milli\meter} CMOS production line including a BEoL. Charge sensing and charge noise measurements at $\approx\SI{20}{\milli\kelvin}$ reveal state-of-the-art characteristics.

\section{Device}  % Pascal
\begin{figure}[htb]
    \centering
    \includegraphics[width=.8\linewidth]{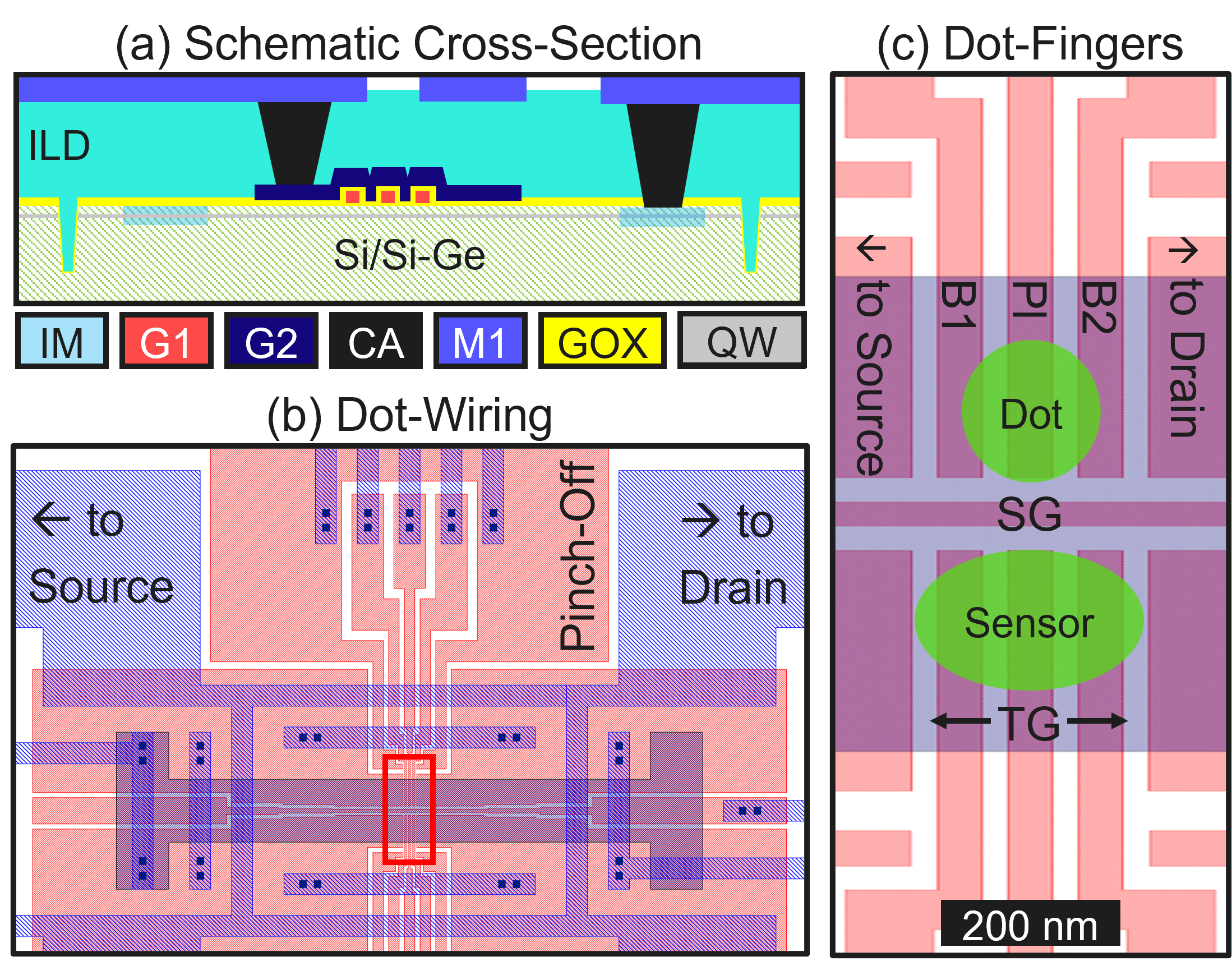}
    \caption{Schematic device cross-section \textbf{(a)} together with the device design labeled with functional names \textbf{(b)} and a zoomed in version for the dot region \textbf{(c)}. The implant regions are labeled as \textbf{IM}, the two different TiN gate layers as \textbf{G1} resp. \textbf{G2}, the gate oxides underneath as \textbf{GOX}. The dots are contacted via leads induced by the wiring layer \textbf{M1}, which is also used for wiring and contacting the gates by tungsten vias labeled as \textbf{CA}. The nominal distance between the two QDs is about 100 nm.}
    \label{fig:layout}
\end{figure}

As substrate a Si/Si--Ge heterostructure with a \SI{10}{\nano\meter} thin, tensile-strained Si layer (\Cref{fig:layout}(a)) is used, in which QDs can be electrostatically formed by tuning gate voltages.
The devices' symmetric gate layout (\Cref{fig:layout}) is designed to benchmark charge noise \cite{Yoneda2018, Struck2020} and valley-splitting \cite{esposti2024low, Volmer_2024} of QDs, both important performance metrics for the heterostructure quality. Of the two interchangeable QDs formed, one can be used as a charge sensing dot to determine the electron occupation of the other.
The QDs are confined laterally by electrostatic potentials from the gate electrodes in the layer \textbf{G1}, as shown in \Cref{fig:layout}(c). The QD is formed by an accumulating plunger gate (\textbf{Pl}), two barrier gates (\textbf{B1, B2}) controlling the coupling to leads, and a gate separating both QDs (\textbf{SG}) from each other. A second gate layer \textbf{G2} acts as top-gate (\textbf{TG}) to globally tune the 2DEG accumulation. The pinch-off gates in G1 are intended to prevent parasitic current paths.
The gates are contacted by tungsten vias (\textbf{CA}) from the separately defined wiring layer \textbf{M1} on top of the interlayer dielectric (\textbf{ILD}).
The implanted areas (\textbf{IM}) of $\approx\SI{2.5e3}{\micro\meter\squared}$ used as ohmic contacts are defined $\approx\SI{60}{\micro\meter}$ away from the active QD region. The wiring layer is used to accumulate charge carriers between the contacts and the active region, inducing reservoirs in the immediate vicinity of the dots to ensure a small stray capacitance.

\begin{figure}[htb]
    \centering
    \includegraphics[width=.8\linewidth]{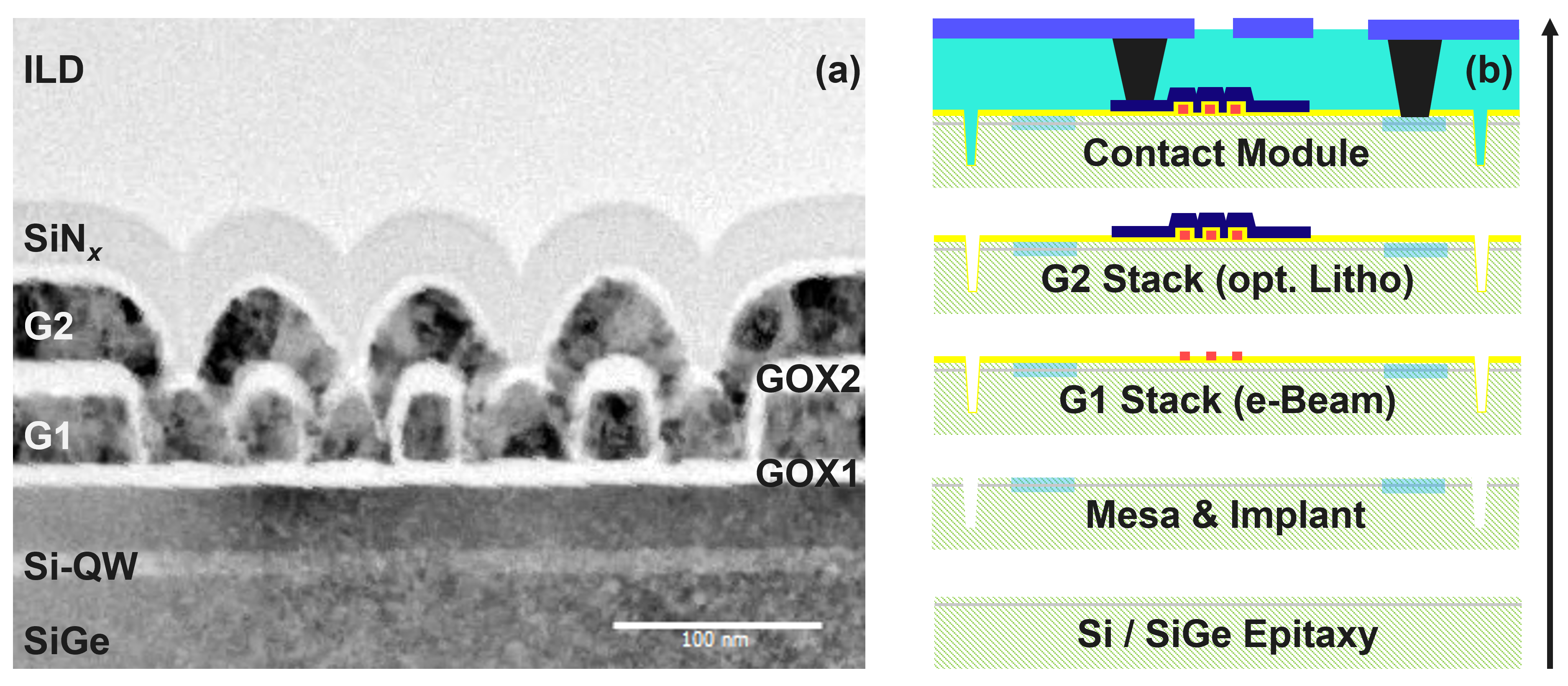}
    \caption{\textbf{(a)} TEM cross-section of an exemplary device \& \textbf{(b)} schematic overview of the relevant process steps, colored as in \Cref{fig:layout}. The TiN metal gates are labeled by their defining layers \textbf{G1} and \textbf{G2}, the underlying PECVD gate oxides as \textbf{GOX1} and \textbf{GOX2} respectively. The etch-stop nitride (\textbf{SiN$_x$}) and interlayer dielectric (\textbf{ILD}) layers lying on top are necessary for the contact module.}
    \label{fig:tem-xs}
\end{figure}

\Cref{fig:tem-xs} shows a transmission electron microscopy (TEM) of an exemplary QD device comparable to the measured ones and a schematic overview of the process flow.

% Heterostructure
The Si/Si--Ge heterostructure was epitaxially grown using CVD with natural isotopic composition on \SI{200}{\milli\meter} $\langle 100 \rangle$ Si wafers and was purchased from Lawrence Semiconductor Research Laboratory Inc.
The barrier underneath the Si quantum well consists of a several \si{\micro\meter} thick linearly graded Si$_{1-x}$Ge$_{x}$ buffer, used for relaxing the Si--Ge, and a several hundred \si{\nano\meter} thick Si$_{0.7}$Ge$_{0.3}$ barrier on top. The strained Si quantum well of \SI{10}{\nano\meter} thickness is covered by a \SI{30}{\nano\meter} thick Si$_{0.7}$Ge$_{0.3}$ spacer layer. It is terminated with a Si cap layer, thinned to roughly \SI{2}{\nano\meter} by the subsequent gate stack processing. These parameters are similar to those commonly used \cite{Neyens2024,esposti2024low,Paqueletwuetz2023}.
% Gate Oxides

The \SI{10}{\nano\meter} thick SiO$_2$ oxide underneath both G1 and G2 is deposited through low temperature ($\approx$\,\SI{400}{\celsius}) plasma enhanced CVD. This SiO$_2$ acts as a good dielectric insulator with low leakage and trap density, down to \SI{3e10}{\per\centi\meter\squared\per\electronvolt}. It thus offers an attractive alternative to aluminum oxides -- with measured interface trap densities in the range of \SIrange{3e11}{1E12}{\per\centi\meter\squared\per\electronvolt} (not shown) -- or other high-$\kappa$ dielectrics deposited by Atomic Layer Deposition, which are generally considered to have a lower interface quality \cite{Wallace2003}.
% Gate Metal
The oxide layers are covered with a \SI{30}{\nano\meter} thick sputtered TiN layer acting as the gate metal for G1 and G2. Only the first fine-patterned gate layer G1 is exposed by an industrially compatible, high-throughput electron-beam lithography (EBL) tool. This is the most effective approach for prototyping and low-volume production \cite{Brackmann2023} on \SI{200}{\milli\meter} without compromising device quality. The second self-aligned gate layer G2 is patterned by optical lithography. The transfer of the photo-resist pattern into the TiN layer after exposure is carried out by dry-etch processes followed by a wet gate stack clean, which removes all metal free gate oxides for subsequent processes. This ensures high-quality oxides without any memory of previous processes introducing defects, charges, or contamination. 
% Contact Module
Subsequently, we process the contact module to contact the double gate stack using a standard \SI{90}{\nano\meter} BEoL. 

\begin{figure}[htb]
    \centering
    \includegraphics[width=\linewidth]{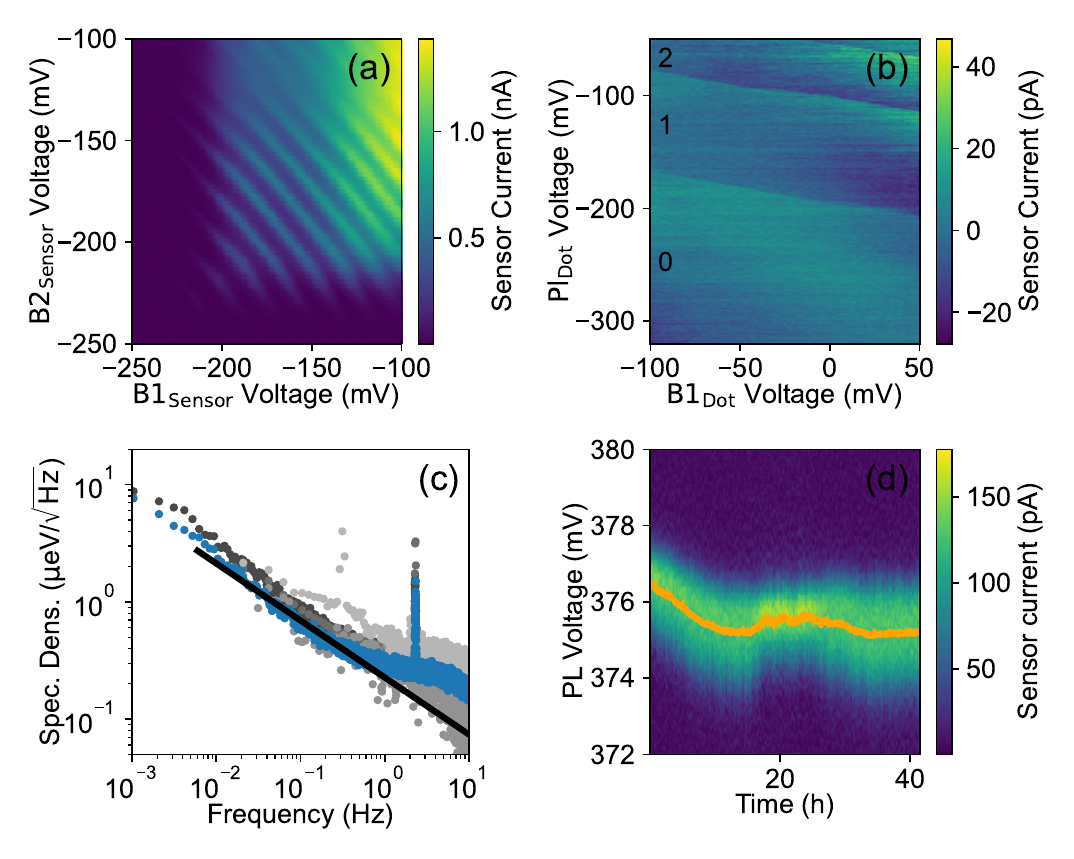}
    \caption{(a) Barrier-Barrier scan of SET with pristine Coulomb oscillations. (b) Charge stability diagram of dot on side 1 measured with side 2 of the device operated as SET. Black numbers indicate the dot occupancy. The sensor background is subtracted to enhance visibility. In (a) and (b), a source-drain bias of \SI{0.1}{\milli\volt} was applied by a lock-in amplifier (c) Noise spectral density, extracted by Welch's method from time traces of the sensor current at the flank of an Coulomb oscillation. The Source-Drain-Bias was increased to \SI{300}{\micro \volt} for these time traces. The black line is a representative fit 
    with $\alpha$=-0.97 to the blue curve. Grey data points correspond to equivalent measurements at different Coulomb peaks. (d) Tracking of a Coulomb peak over $\approx$ \SI{40}{\hour} by continuous PL voltage sweeps. The peak position, extracted by fitting a Gaussian function to the individual traces, is indicated in orange.}
    \label{fig3}
\end{figure}

%Section: Measurement results
\section{Millikelvin Characterization}
We tested more than 10 devices at cryogenic temperatures and provide a detailed report on the characteristics of the first fully functional device, which we measured in a dilution refrigerator with a base temperature of \SI{20}{\milli\kelvin}. Several other devices gave a similar picture regarding disorder and drift of the quantum dots.
By applying wafer-scale cryo-characterization of similar QD devices fabricated in the same way, a leakage ($R>\SI{100}{\mega\ohm}$), accumulation ($S>\SI{1}{\micro S}$) and gate functionality yield $\ge\SI{90}{\percent}$ have been achieved.

We start the characterization by finding a set of gate voltages that allows the formation of a single-electron transistor (SET) for charge sensing and an adjacent few electron quantum dot, a process called tuning.
To this end, we increase the voltages of accumulation gates in G1 and M1, of the topgate TG and the finger gates B1, Pl and B2 to accumulate charge carriers in the quantum well, thus forming a conductive channel. Once a sensor current is observable, we reduce the barrier gates' voltages to raise tunneling barriers and form a dot close to Pl.
Using small adjustments to Pl, TG and SG, we then form a SET featuring clean Coulomb oscillations as visible in \Cref{fig3}(a). The straightness of the Coulomb lines and absence of resonance features attest low disorder.
The symmetry of the device allows us to tune its second dot in analogy to the first one. Note that tuning of the second dot usually requires a re-adjustment of the SET gate's voltage due to the cross-coupling of the dot and the SET gates. Subsequently, B2 of the dot is closed, which facilitates reaching the few electron regime. Ramping B1's and Pl's voltages to control tunnel barrier and dot potential, respectively, we measure the charge stability diagram (\Cref{fig3}(b)). Steps reflecting changes of the dot occupancy are clearly visible, and the electron filling can be controlled down to the last electron. 
Functional devices 
%that are not subject to the previously mentioned fabrication issues 
show good tunability and can be brought into the single-electron regime within a few hours. Furthermore, variations between multiple thermal cycles for the same device are minor.

As dephasing due to nuclear spins can be reduced significantly by using a Si$^{28}$ enriched quantum well, charge noise is the most relevant technological parameter determining qubit coherence \cite{Yoneda2018,Struck2020,Elsayed2024}. It is predominantly influenced by the quality of the gate stack, such as defects in the gate oxide and the oxide-substrate interface \cite{Kepa2023}.
To benchmark long term stability and noise characteristics of the device, we record several time traces of the SET at operation points on the flanks of multiple Coulomb oscillations. We determine the spectral density $\sqrt{S} = \frac{\alpha}{dI/dV}\times \sqrt{S_{i}}$ of the electrochemical potential of the quantum dot with $dI/dV$ the slope of the respective Coulomb peak at the operation point, an averaged lever arm $\alpha$ of \SI{0.076 \pm 0.011}{\electronvolt\per\volt} extracted from Coulomb diamonds and the power spectral density $S_{i}$ of the current measurement as calculated from the time traces using Welch's method. The results are shown in \Cref{fig3}(c). The measured lever arm lies within the range of similarly investigated values on related Si/SiGe heterostructures from other groups \cite{Neyens2024,Paqueletwuetz2023,Connors2019}. Below \SI{0.1}{\hertz} we observe a $f^{\alpha/2}$ behavior with an average of $\alpha = \SI{-0.99 \pm 0.04}{}$. The kink at frequencies around \SI{0.1}{\hertz} is likely caused by the setup's noise background. Extrapolation of the $f^{\alpha/2}$ behavior at lower frequencies to \SI{1}{\hertz} yields an average noise spectral density of \SI{0.27 \pm 0.05}{\micro\electronvolt\per\hertz\tothe{1/2}}, which is among the lowest values found in devices from academic fabrication \cite{Struck2020,esposti2024low,Paqueletwuetz2023,Connors2019}. It is comparable to the value resulting from an extensive gate stack optimization reported on Si/Si--Ge \cite{Paqueletwuetz2023} with an average of \SI{0.29 \pm 0.02}{\micro\electronvolt\per\hertz\tothe{1/2}}.
%In Si-MOS, (i.e., not utilizing a heterostructure), an average of \SI{0.61}{\micro\electronvolt\per\hertz\tothe{1/2}} has been reported recently \cite{Elsayed2024}.
We obtained similar results in a second cooldown of the same device in a different setup.
Coulomb peaks tracked over the course of \SI{40}{\hour} exhibit a very low drift with shifts of less than \SI{2}{\milli\volt}. This is an important test, as electron tunneling to states at the Si-SiO$_2$ interface can lead to strong drift and instability phenomena in some heterostructure-based quantum devices \cite{Huang2014, Ferrero24}.

Another important characteristic of Si/Si--Ge QDs is the energy splitting $E_\mathrm{VS}$ between the lowest conduction band valleys. It depends on local atomistic details of the Si--Ge heterostructure and small $E_\mathrm{VS}$ compromise spin qubit operations \cite{Kawakami2014}. Narrower QWs with slightly higher Ge-content or just Ge-inter-diffusion at the QW interfaces increase $E_\mathrm{VS}$ on average \cite{Klos2024,Paqueletwuetz2023}.
From magneto-spectroscopy \cite{esposti2024low} of the consecutive charging lines for a QD occupation from 0 to 3 electrons \cite{mcjunkin2021}, we deduce $E_{VS} = \SI{50}{\micro\electronvolt}$ for a dot configuration corresponding to a gate-controlled confinement. This comparatively low value is within the range of values found in a valley splitting mapping for a device fabricated from the same batch of heterostructures by an academic liftoff process \cite{Volmer_2024}.

\section{Conclusion}
We have demonstrated few electron QD devices in Si/Si--Ge heterostructures fabricated within an industrial production line including a high-quality, low-temperature grown PECVD SiO$_2$, a BEoL process and CMOS-compatible electron-beam lithography. We evaluated the device performance in terms of the quality of Coulomb oscillations of a single-electron transistor, charge noise, single electron detection and valley-splitting. Although the study focuses on few exemplary devices and is to be verified by larger statistics, our findings reveal that the industrial etching process and the limited thermal fabrication budget does not compromise the device performance. 
On the contrary, the charge-noise and electrostatic disorder in the Si quantum well is remarkably low and a correlation to the low trap density of the gate oxides could be speculated. 

We expect that our process is easily adaptable to the realization of Ge quantum dots, which are advantageous regarding all-electrical spin control and lithography requirements but are much more sensitive to charge noise \cite{Scappucci2021}.
On either platform, the expected high level of yield and reliability, in contrast to an academic metal lift-off process, and a multilayer BEoL are a prerequisite for two-dimensional quantum processor architectures such as the one proposed by K{\"u}nne \textit{et al.} \cite{Kunne2024}.
% Comment towards shuttling?
%In the near future, conveyor mode shuttling on similarly fabricated devices will be presented.

\section*{Acknowledgements} \label{sec:acknowledgements}
This work was partially funded by the German Federal Ministry of Education and Research, reference numbers 13N15652, 13N14778, 13N15658, Project QUASAR and Project SiQuBus; the German Research Foundation within the project 289786932 (BO 3140/4-2) and the European Union's Horizon research and innovation program, grant agreement No. 951852, Project QLSI.

\bibliography{references}

%\appendix*
%\input{content/appendix}
\end{document}